\documentclass[leqno]{article}
\usepackage[letterpaper]{geometry}

\usepackage{cite}
\usepackage{amsmath,amssymb,amsfonts}
\usepackage{amsthm}
\usepackage{graphicx}
\usepackage{textcomp}
\usepackage{xcolor}
\usepackage{booktabs}
\usepackage{xspace}
\def\BibTeX{{\rm B\kern-.05em{\sc i\kern-.025em b}\kern-.08em
    T\kern-.1667em\lower.7ex\hbox{E}\kern-.125emX}}

 \usepackage[show]{chato-notes}

\usepackage{subfigure}

\newtheorem{problem}{Problem}[section]

\newtheorem{theorem}{Theorem}[section]
\newtheorem{lemma}[theorem]{Lemma}

\newcommand{\bigO}{\ensuremath{\mathcal{O}}\xspace}
\newcommand{\tbigO}{\ensuremath{\tilde{\mathcal{O}}}\xspace}

\newcommand{\ceil}[1]{\lceil #1 \rceil}

\newcommand{\N}{\ensuremath{\mathbb{N}}\xspace}
\newcommand{\E}{\ensuremath{\mathbb{E}}\xspace}

\usepackage{algorithm, algpseudocode}
\renewcommand{\Return}{\State \textbf{return }}
\newcommand{\Continue}{\State \textbf{continue }}
\newcommand{\argmin}{\arg\min\limits}

\newcommand{\subgraph}[1]{\subseteq_{#1}}
\renewcommand{\algref}[1]{{Algorithm~\ref{alg:#1}}}
\newcommand{\tabref}[1]{{Table~\ref{tab:#1}}}
\newcommand{\figref}[1]{{Figure~\ref{fig:#1}}}
\newcommand{\block}[1]{\smallskip\noindent{\textbf{#1}}}

\newcommand{\mh}{{\small MH}\xspace}
\newcommand{\mcmc}{{\small MCMC}\xspace}
\newcommand{\psrw}{{\small PSRW}\xspace}
\newcommand{\mcmcsampling}{{{\small MCMC\-S}}{\sc am\-pling}\xspace}
\newcommand{\mixingkmcmc}{\ensuremath{t_k^{\mathrm{MCMC}}}\xspace}
\newcommand{\mixingkpsrw}{\ensuremath{t_k^{\mathrm{PSRW}}}\xspace}
\newcommand{\rss}{{\small RSS}\xspace}
\newcommand{\rssopt}{{\small RSS}{+}\xspace}

\newcommand{\squishlist}{
 \begin{list}{$\bullet$}
  {  \setlength{\itemsep}{0pt}
     \setlength{\parsep}{3pt}
     \setlength{\topsep}{3pt}
     \setlength{\partopsep}{0pt}
     \setlength{\leftmargin}{2em}
     \setlength{\labelwidth}{1.5em}
     \setlength{\labelsep}{0.5em}
} }
\newcommand{\squishlisttight}{
 \begin{list}{$\bullet$}
  { \setlength{\itemsep}{0pt}
    \setlength{\parsep}{0pt}
    \setlength{\topsep}{0pt}
    \setlength{\partopsep}{0pt}
    \setlength{\leftmargin}{2em}
    \setlength{\labelwidth}{1.5em}
    \setlength{\labelsep}{0.5em}
} }

\newcommand{\squishdesc}{
 \begin{list}{}
  {  \setlength{\itemsep}{0pt}
     \setlength{\parsep}{3pt}
     \setlength{\topsep}{3pt}
     \setlength{\partopsep}{0pt}
     \setlength{\leftmargin}{1em}
     \setlength{\labelwidth}{1.5em}
     \setlength{\labelsep}{0.5em}
} }

\newcommand{\squishend}{
  \end{list}
}

\newcommand*{\nvector}[1]{\mathrm{#1}}
\newcommand*{\nmatrix}[1]{\uppercase{\mathbf{#1}}}

\begin{document}


\title{{\Large Improved mixing time for $k$-subgraph sampling}\thanks{This
work was supported
by the Academy of Finland project 
``Active knowledge discovery in graphs (AGRA)'' (313927), 
the EC H2020 RIA project ``SoBigData++'' (871042), 
and the Wallenberg AI, Autonomous Systems and Software Program (WASP)
funded by Knut and Alice Wallenberg Foundation.}}
\author{Ryuta Matsuno\thanks{Tokyo Institute of Technology, Japan. 
This work was done while the author was visiting Aalto University.}
\and Aristides Gionis\thanks{KTH Royal Institute
of Technology, Sweden, and Aalto University, Finland.}}

\date{}

\maketitle








\begin{abstract} 
Understanding the local structure of a graph provides valuable insights 
about the underlying phenomena from which the graph has originated.
Sampling and examining $k$-sub\-graphs is a widely used approach to understand the local struc\-ture of a graph. 
In this paper, we study the problem of sampling uniformly $k$-subgraphs from a given graph. 
We analyse a few different Markov chain Monte Carlo (\mcmc) approaches, 
and obtain analytical results on their mixing times,
which improve significantly the state of the art. 
In particular, we improve the bound on the mixing times of the standard \mcmc approach, 
and the state-of-the-art \mcmc sampling method \psrw, using the {\em canonical-paths} argument. 
In addition, we propose a novel sampling method, 
which we call \emph{recursive subgraph sampling} \rss, and its optimized variant \rssopt. 
The proposed methods, \rss and \rssopt, are provably faster than the existing approaches.
We conduct experiments and verify the uniformity of samples and the time efficiency of \rss and \rssopt.

\end{abstract}

\section{Introduction}

Graphs are used to model complex real-world data in a wide range of domains, 
such as, sociology, biology, ecology, transportation, telecommunications, and more. 
Understanding the structural properties of graphs, at different levels of granularity, 
provides valuable insights about the underlying phenomena and processes that generate 
the corresponding graph data.
A compelling approach to explore the structural properties of a graph, or a collection of graphs, 
at a fine scale, is to extract information about small-size subgraphs related to their connectivity patterns, 
interactions, and other features of 
interest~\cite{chen2006nemofinder,milo2002network,kunegis2009slashdot,toivonen2006model,zhao2011empirical}.
For instance, the high ratio of closed triangles observed in social networks has been considered
a manifestation of social affinity observed in human society and which 
leads to forming tightly-knit groups.
As a concrete example, it has been found that the ratio of closed triangles is higher in facebook, 
which is primarily an online social network, than in twitter, which is used as a
platform for news dissemination~\cite{kwak2010twitter}.

More interesting structural properties and hidden patterns in the graph data can be revealed by examining larger subgraphs,  e.g., subgraphs of size $k$, or $k$-subgraphs.
Unfortunately, the number of $k$-subgraphs in a given graph increases exponentially with $k$, 
and enumerating all possible $k$-subgraphs becomes prohibitive. 
To address this challenge one usually resorts to sampling. 
To make the sampling idea viable requires obtaining a representative subset of the set of all $k$-subgraphs, 
or equivalently, sampling $k$-subgraphs \emph{uniformly at random}, 
which is a challenge by itself.
As a result, the problem of uniform sampling $k$-subgraphs, 
has been extensively studied in data mining, statistics, and theoretical computer science~\cite{ahmed2015efficient,bhuiyan2012guise,bordino2008mining,bressan,han2016waddling,jha2015path,pinar2017escape,wang2014efficiently,wang2017moss}.

In this paper, we study the problem of sampling uniformly at random 
$k$-subgraphs from a given input graph.
Among the different methodologies that have been proposed, 
we focus on the Markov chain Monte Carlo (\mcmc) 
approach~\cite{metropolis1953equation}, 
and in particular on the Metropolis-Hastings algorithm (\mh)~\cite{Hastings}.
The high-level idea is to sample from the stationary distribution of
a Markov chain, whose set of states is the set of $k$-subgraphs,
by performing a random walk. 
The \mh algorithm~\cite{Hastings} is used to ensure that the stationary distribution, and thus, the sampling, is uniform.
An important theoretical question is to upper bound the  \emph{mixing time} of the random walk, 
which is the time needed for the empirical sampling distribution to be close enough to the stationary distribution.

We present improved results for the mixing time of Markov chains designed for uniform sampling of $k$-subgraphs.
Our starting point is the recent work of Bressan et al.~\cite{bressan}, 
who analyze a \mcmc method 
and show that it mixes
in time $\tbigO((k!)^2\Delta^{2k}|V|^2)$, 
where $|V|$ is the number of nodes in the input graph,
$\Delta$ is a maximum node degree,
and $\tbigO(\cdot)$ is used to suppress logarithmic and other 
lower-term factors.

Our first result is to analyze the Markov chain with the \mh algorithm using the technique of \emph{canonical paths}~\cite{mix}
and obtain an upper bound on the mixing time of 
$\tbigO(k!\Delta^{k}(D+k)|V|)$, where $D$ is a diameter of the graph. 
Our bound is a significant improvement of the bound of Bressan et al.~\cite{bressan}.

Next, we proceed to improve this bound even further, 
by introducing a novel Markov chain to perform the random walk. 
In particular, we propose a technique based on 
\emph{recursive subgraph sampling} (\rss),
and a further improvement called \rssopt,  
which exploits the fact that we can easily sample $2$-subgraph uniformly at random in time $\bigO(1)$: sampling a $2$-subgraph is just sampling an edge on the input graph. In turn, this gives us a way to sample a $3$-subgraph uniformly at random by applying the \mh approach. The idea can be applied recursively for all $k$.
The complexity of the \rss scheme is 
$\tbigO(c^{k-3}(k!)^2 ((k-1)!)^2 \Delta^{k-3})$, 
where $c$ is a constant. 
Although this bound is still large, 
assymptotically is a big improvement compared to the previous work,
and to our \mcmc bound.
Furthermore, in practice, $k$ is typically taken to be a small constant, i.e., $|V| \gg k$.

We experimentally evaluate our methods for $k=3,4,5$
and we verify the superiority of the \rss scheme
against the \mcmc method, and the state-of-the-art \psrw algorithm~\cite{psrw}.
We also evaluate \rss and \rssopt for values of $k$ up to 10;
we present this in the Appendix.
As it has been observed by other researchers, 
we confirm that the theoretical 
bounds are overly pessimistic, 
and in practice it suffices to run the random walk for a significantly smaller number of steps.

\vspace{1mm}
Our key contributions are as follows:
\squishlist
\item
We apply  the technique of canonical paths~\cite{mix} 
to obtain a bound on the mixing time of the standard \mcmc\ method, 
which is a significant improvement over the state of the art.
\item 
We propose novel $k$-subgraph sampling algorithms, \rss and \rssopt, 
whose computational costs further improve the 
mixing time of \mcmc sampling. 
\item 
We obtain a bound on the mixing time of the \psrw method~\cite{psrw}, 
which was left open by the authors. 
\item 
We conduct an experimental evaluation and show that \rss and \rssopt are significantly faster 
compar\-ed to \mcmc\ sampling and \psrw.
\squishend

The rest of this paper is organized as follows.
We start by reviewing the related work 
in Section~\ref{section:related}.
Our algorithms, analysis, and main results are presented 
in Section~\ref{section:sampling}.
In Section~\ref{section:experiments} we discuss our experimental evaluation,
and finally, Section~\ref{section:conclusion} is a short conclusion.

The proofs of our lemmas and claims can be found in Appendix~\ref{section:proofs}, 
while additional experimental results are presented in Appendix~\ref{section:additional-experiments}. 

\section{Related work}
\label{section:related}

Sampling $k$-subgraphs uniformly at random is computationally expensive. 
The number of possible $k$-subgraphs of $G=(V,E)$ is $\binom{|V|}{k} = \bigO(|V|^k)$. 
Enumerating all of them is intractable. 
Hence, the approximation of sampling has been studied \cite{lu, psrw, bressan, guise}.

A standard approach is to apply Markov chain Monte Carlo (\mcmc) sampling with the Metropolis-Hastings (\mh) technique \cite{lu, psrw, bressan, guise}. This approach performs a random walk on a graph whose nodes are all $k$-sub\-graphs of an input graph $G=(V,E)$, and two $k$-subgraphs are adjacent if they differ by one node. The graph of $k$-subgraphs is defined as the \textit{$k$-state graph} in the next section. 
By performing random walk on the $k$-state graph we 
can obtain a uniform sample of $k$-sub\-graphs. 
Bressan et al.~\cite{bressan} study the con\-ductance of the $k$-state graph and show that it increases expo\-nen\-tial\-ly, which directly implies that the mixing time of a simple random walks on it also increases exponentially; the upper bound of the mixing time is $\tbigO((k!)^2|V|^2\Delta^{2k})$, where $|V|$ is the number of nodes and $\Delta$ is the maximum degree of the given graph~$G$. 
They also show that even when the given graph has low conductance, the mixing time can be exponential. 

Wang et al.~\cite{psrw} notice that a $k$-subgraph sample can be obtained by sampling an edge from the graph of $(k-1)$-subgraphs, i.e., the $(k-1)$-state graph. This method is named {\em pairwise subgraph random walk}~(\psrw). They prove that \psrw samples a $k$-subgraph uni\-formly at random. Since a random walk on the $(k-1)$-state graph has faster mixing time than on the $k$-state graph, \psrw is more efficient than the standard \mcmc approach. It is, however, still~exponential.

As another approach, Bressan et al.~\cite{bressan} propose a sampling algorithm that uses the color-coding technique \cite{naga1995cc}. The computational cost of their method is $\bigO(c^k|E|)$. 
In this paper, we focus on Markov chain approaches, and we analyse the mixing time of $k$-state graphs. Thus, we exclude this approach from our comparisons for the sake of consistency.

\section{Subgraph sampling}
\label{section:sampling}

\subsection{Terminology and problem definition}

We start with an undirected graph $G = (V,E)$.
Let $H = (V_H,E_H)$ be a \emph{connected $k$-subgraph} of $G$, that is, 
$H$ is a connected vertex-induced subgraph of $G$ containing exactly $k$ nodes. 
More precisely, $V_H \subseteq V$, $|V_H| = k$, $E_H = \{ (u,v) \in E \mid u, v \in V_H \}$, and $H$ is connected.
We use the notation $H \subgraph{k} G$ to denote that $H$ is a 
connected $k$-subgraph of $G$.

Let $V^{(k)}$ be a set of all connected $k$-subgraphs of~$G$, i.e., 
$V^{(k)} = \{ H \subgraph{k} G \}$. 
We construct a graph $G^{(k)}$ whose node set is $V^{(k)}$. 
In the graph $G^{(k)}$, 
two nodes $H = (V_{H},E_{H})$ and $F = (V_{F},E_{F})$ are adjacent if and only if 
the sets $V_{H}$ and $V_{F}$ differ by exactly one node. 
Hence, the edge set $E^{(k)}$ of the constructed graph $G^{(k)}$ is:
\begin{align}
\begin{small}
E^{(k)} = \{(H,F) \mid  H,F \in V^{(k)} \text{ and } | V_{H} \cap V_{F}| = k-1 \}. \notag
\end{small}
\end{align}
The graph $G^{(k)}=(V^{(k)},E^{(k)})$ defined above is called 
the \emph{$k$-state graph} of $G$. 
Note that $G=(V,E)$ can also be seen as $G^{(1)} = (V^{(1)},E^{(1)})$ for the case of $k=1$.

We denote by $\Delta$ the maximum degree of a node in $G$, and 
by $\Delta_k$ the maximum degree of a node in $G^{(k)}$. 
We denote by $d(u)$ the degree of node $u$ in the graph that $u$ belongs;
e.g., if $H \subgraph{k} G$, then $d(H)$ denotes the degree of $H$ in the $k$-state graph $G^{(k)}$.
Note that Bressan et al.~\cite{bressan} study $k$-state graphs and upper bound the maximum degree of $k$-subgraph by $\Delta_k \leq k \Delta$. 
They also give an upper bound of the number $|V^{(k)}|$ of $k$-subgraphs by $|V^{(k)}| \leq (k-1)!\Delta^{k-1}|V|$.

The problem we consider in this paper is to sample uniformly at random
a node from the graph $G^{(k)}$, given a graph $G$ and an integer $k$.
More formally:
\begin{problem}[Uniform $k$-subgraph sampling]
\label{problem:k-subgraph-sampling}
Given a graph $G = (V,E)$ and a number $k \in \N$, with $1 < k < |V|$, 
sample a connected $k$-subgraph $H \subgraph{k} G$ uni\-form\-ly at random.
\end{problem}

\subsection{Overview}
Before presenting the proposed solution for \textsc{Problem}~\ref{problem:k-subgraph-sampling},
we review the standard Markov Chain Monte Carlo (\mcmc) approach, 
and introduce concepts needed in our analysis.


\block{\mcmc method.}
The \mcmc method is used to obtain a sample from a desired distribution by designing a Markov chain whose stationary distribution corresponds to the desired distribution. 
Let $\Omega = \{0,\ldots, m-1\}$ be a state space, 
and $p(u,v)$ be the transition probability between states $u,v\in\Omega$, 
also represented as a matrix of transition probabilities
$\nmatrix{P} \in [0,1]^{m \times m}$, with $\nmatrix{P}_{u,v} = p(u,v)$.
Starting from $x \in \Omega$, 
the probability that a random walk visits $y \in \Omega$ in exactly~$t$ steps 
is given by $(\nvector{e_x} \nmatrix{P}^t)_y$, 
where $\nvector{e_x}$ is a unit row vector having 1 in $x$-th coordinate.
An ergodic Markov chain has a stationary distribution $\pi \in [0,1]^m $, given by $\pi = \pi \nmatrix{P}$. 
Hence, by conducting a sufficiently long random walk on the chain
we can obtain a sample from the distribution~$\pi$.

\block{Metropolis-Hastings algorithm.}
The Metropolis-Hastings algorithm (\mh) \cite{Hastings} is a standard technique 
to convert a stationary distribution $\pi$ of a Markov chain to a desired stationary distribution $\pi'$. 
It adds one step in \mcmc sampling:
a transition from $x\in \Omega$ to $y \in \Omega$
is accepted with probability $\min \left\{1,\frac{\pi'_y/\pi_y}{\pi'_x/\pi_x}\right\}$, 
otherwise, the walk stays at $x$.
The resulting random walk has stationary distribution $\pi'$.

\block{Mixing time of \mcmc.}
Mixing time provides a measure of efficiency of the sampling method
by quantifying how fast the sampling distribution $\nvector{e_x}\nmatrix{P}^t$, starting at state $x$,  
approaches the stationary distribution~$\pi$ \cite{mix, mixbook}.
The mixing time $\tau(\varepsilon)$ is defined as the minimum number of random-walk steps 
required to achieve quality of approximation $\varepsilon$. In particular, 
\begin{align*}
\tau_x(\varepsilon) =
\arg & \min_{t \in \N} 
\left\{  
\frac{1}{2} \sum_{y\in \Omega} |(\nvector{e_x}\nmatrix{P}^{\ell})_y - \pi_y| \leq \varepsilon, \text{ for all } \ell \geq t 
\right\},\\
\text{and }~
\tau(\varepsilon)  = & \max_{x \in \Omega} \tau_x(\varepsilon).
\end{align*}
%

\block{Canonical paths}~\cite{mix}. 
This term refers to a proof technique used
to upper bound the mixing time of a Markov chain.
Given a Markov chain with state space~$\Omega$, 
we define an underlying directed graph $(\Omega,E_M)$,
where $E_M$ is a set of directed edges between states in $\Omega$
with positive transition probability, i.e., 
$E_M = \{(u,v)\in \Omega \times \Omega \mid p(u,v) > 0 \}$.
A canonical path $\gamma_{xy}$ is a path from $x \in \Omega$ to 
$y \in \Omega$ on the graph $(\Omega,E_M)$. 
A set of canonical paths $\Gamma$ consists of canonical paths 
for each ordered pair of distinct states $x,y \in \Omega$.
An upper bound on the mixing time can be calculated as follows~\cite[Proposition 12.1]{mixbook}:
\begin{align}
  \label{equation:mixing-rho}
  & \tau_x(\varepsilon)
  \leq \overline\rho\,(\ln {\pi_x}^{-1} + \ln \varepsilon^{-1}),
\\
\nonumber 
\text{where }~
 & \overline  \rho 
  = \max_{(u,v) \in E_M} \frac{1}{Q(u,v)} \sum_{\gamma_{xy} \in \Gamma \wedge \gamma_{xy} \ni (u,v)} \pi_x \pi_y |\gamma_{xy}|,
\end{align}
$Q(u,v) = \pi_u p(u,v) = \pi_v p(v,u)$, and $|\gamma_{xy}|$ is the length of the path $\gamma_{xy}$. 
The tightness of the upper bound depends on the choice of canonical paths. 
Intuitively, we want to select canonical paths so that no single edge is used by too many paths.
More details can be found in the excellent book of Jerrum and Sinclair~\cite{mixbook}.

\subsection{Markov Chain Monte Carlo (\mcmc) approach}

A simple solution to $k$-subgraph sampling
(Problem~\ref{problem:k-subgraph-sampling})
is to apply the \mcmc and \mh methods discussed above. 
The method is shown in \algref{MCMC}, and we refer to it as \mcmcsampling.
The main observation is that the stationary distribution of a random walk in an 
undirected graph is proportional to the node degrees, 
thus, adding the acceptance probability step in line~\ref{line:mh}, 
according to \mh, 
leads to uniform sampling. 
Note that the condition in line~\ref{line:self-loop} adds a $\frac{1}{2}$-probability self-loop 
to ensure non-periodicity.

\begin{algorithm}[t]
\begin{small}
  \caption{\mcmcsampling}
  \label{alg:MCMC}
  \begin{algorithmic}[1]
    \Require Graph $G = (V,E)$, subgraph size $k$, error $\varepsilon>0$
    \Ensure $H \subgraph{k} G$ sampled from $V^{(k)}$ uniformly at random
      \State $v_c \gets$ arbitrary node in $G^{(k)}$
      \State $N_c \gets$ neighbor nodes of $v_c$ in $G^{(k)}$ 
      \For{$\ceil{\mixingkmcmc(\varepsilon)}$ times}
        \If{$\mathit{random}(0,1) < \frac{1}{2}$} \label{line:self-loop}
          \State $v_n \gets$ randomly selected node from $N_c$
          \State $N_n \gets$ neighbor nodes of $v_n$ in $G^{(k)}$ 
          \If{$\mathit{random}(0,1) < \min \left\{ 1,\frac{d(v_c)}{d(v_n)}\right\}$} \label{line:mh}
            \State $v_c \gets v_n, N_c \gets N_n$
          \EndIf
        \EndIf
      \EndFor
      \Return $v_c$
  \end{algorithmic}
\end{small}  
\end{algorithm}

To bound the mixing time of \mcmcsampling,
we apply the canonical-paths technique. 
First note that the Markov chain of \mcmcsampling is on $G^{(k)} = (V^{(k)},E^{(k)})$. 
We choose a canonical path $\gamma_{xy}$ to be one of the shortest paths from $x$ to $y$ on $G^{(k)}$. 
The length of the path $|\gamma_{xy}|$ is bounded by the diameter of $G^{(k)}$, 
which in turn can be bounded using the following Lemma.

\begin{lemma}[Diameter of $k$-state graph $G^{(k)}$]
\label{lem:diameter}
The diameter of $k$-state graph $G^{(k)}$ is at most $(D+k-1)$, where $D$ is a diameter of $G$.
\end{lemma}

On the other hand, it is possible to construct problem instances
in which the graph $G^{(k)}$ has a \emph{bottleneck edge}, 
i.e., $V^{(k)}$ consists of two parts which are connected by just one edge.
Then,
\begin{small}
\begin{eqnarray*}
\overline\rho
&=& \max_{(u,v) \in E_M} \frac{1}{Q(u,v)} \sum_{\gamma_{xy} \in \Gamma \wedge \gamma_{xy} \ni (u,v)} \pi_x \pi_y |\gamma_{xy}| \\
&\leq& \max_{(u,v) \in E_M} 2\Delta_k |V^{(k)}| \sum_{\gamma_{xy} \in \Gamma \wedge \gamma_{xy} \ni (u,v)} \frac{1}{|V^{(k)}|^2} (D+k-1) \\
&\leq& 2k\Delta \frac{1}{|V^{(k)}|} (D+k-1) \max_{(u,v) \in E_M}  |\{\gamma \in \Gamma \mid \gamma \ni (u,v) \}| \\
&\leq& 2k\Delta \frac{1}{|V^{(k)}|} (D+k-1) \left(\frac{|V^{(k)}|}{2}\right)^2 \\
&\leq& \frac{1}{2} k! \Delta^{k} (D+k-1)|V|,
\end{eqnarray*}
\vspace{-2mm}
\end{small}

\noindent
where $\pi_x = \frac{1}{|V^{(k)}|}$ for all $x \in V^{(k)}$, $p(u,v) \geq\frac{1}{2\Delta_k}$ for all $(u,v)$, $\Delta_k$ is the maximum degree in $G^{(k)}$, and $\Delta_k \leq k\Delta$ \cite{bressan}.
The largest value for $\max_{(u,v) \in E_M}  |\{\gamma \in \Gamma \mid \gamma \ni (u,v) \}|$ 
on a graph with $|V^{(k)}|$ nodes is achieved when $G^{(k)}$ consists of two parts, 
each of which contains $\frac{|V^{(k)}|}{2}$ nodes, and they are connected by a single edge $(u,v)$. 
Since any path from a state in the one part to a state in the other part goes through $(u,v)$, 
we can bound $|\{\gamma \in \Gamma \mid \gamma \ni (u,v) \}|$ by $\left(\frac{|V^{(k)}|}{2}\right)^2$.
Following Bressan et al., we use the bound $|V^{(k)}| \leq (k-1)! \Delta^{k-1}|V|$. 
An upper bound on the mixing time can now be obtained using Inequality~(\ref{equation:mixing-rho}):
\begin{small}
\begin{eqnarray*}
  \mixingkmcmc(\varepsilon) 
  &\leq& \max_{x \in V^{(k)}} \overline\rho (\ln {\pi_x}^{-1} + \ln \varepsilon^{-1}) \\
  &\leq& \frac{1}{2} k! \Delta^{k} (D+k-1)|V| (k \ln |V| + \ln \varepsilon^{-1}) \\
  & = & \bigO(k!k\Delta^{k} (D+k)|V| \ln |V|).
\end{eqnarray*}
\end{small}

The mixing time gives a bound on the number of random-walk steps required
to obtain one sample.
For the total computational cost of \mcmcsampling,
we also need to consider the cost per random-walk step.
The number of neighbor nodes from a node in $G^{(k)}$ is $\bigO(k^2\Delta)$, 
and it takes $\bigO(k^2)$ to check whether such a neighbor is connected, 
giving a cost of $\bigO(k^4\Delta)$ per random-walk step.
The total cost of \mcmcsampling is 
$\bigO(k!k^5\Delta^{k+1} (D+k)|V| \ln |V|)$.

Based on the analysis so far, we obtain the following results
regarding the mixing time and the computational cost of \mcmcsampling. 

\begin{lemma}[Mixing time] 
\label{lem:mcmcsampling}
The mixing time of the \mcmcsampling algorithm is upper-bounded by 
$\frac{1}{2} k! \Delta^{k} (D+k-1)|V| (k \ln |V| + \ln \varepsilon^{-1})$ 
$= \bigO(k!k\Delta^{k} (D+k)|V| \ln |V|)$.
\end{lemma}
\begin{theorem}[Computational cost] 
\label{thm:mcmcsampling}
The running time of the \mcmcsampling algorithm is
$\bigO(k!k^5\Delta^{k+1} (D+k)|V| \ln |V|)$.
\end{theorem}

\subsection{Recursive subgraph sampling}

\mcmcsampling has provable guarantee on the mixing time, 
however, its complexity is prohibitive.
Thus, we would like to develop an improved sampler with lower complexity.
Next, we develop a \emph{recursive subgraph sampling} (\rss) algorithm, 
which also samples a connected $k$-subgraph from a given graph with uniform probability.
\rss is shown in Algorithms~\ref{alg:rec} and \ref{alg:prop}. 
The main function 
$\Call{Uniform\-Sampling}{G,k,\varepsilon}$ and 
the subroutine
$\Call{Degree\-Prop\-Sampling}{G,k,\varepsilon}$ 
call each other $(k-3)$ times in a recursive manner. 

\begin{algorithm}[t]
\begin{small}
  \caption{Recursive subgraph sampling (RSS)}
  \label{alg:rec}
  \begin{algorithmic}[1]
    \Require Graph $G = (V,E)$, subgraph size $k$, error $\varepsilon>0$
    \Ensure $H \subgraph{k} G$ sampled from $V^{(k)}$ uniformly at random
    \Function{UniformSampling}{$G,k,\varepsilon$}
      \If{$k = 2$}
        \Return a uniformly-sampled edge from $E$
      \EndIf
      \While{{\small TRUE}}
        \State $v \gets $\Call{Degree\-Prop\-Sampling}{$G,k-1,\varepsilon$}
        \State $u \gets $ uniformly-sampled neighbor of $v$ in $G^{(k-1)}$
        \State $H \gets$ $k$-subgraph with nodes in $v$ and $u$
        \State $m \gets $ number of $(k-1)$-subgraphs in $H$
        \If{$\mathit{random}(0,1) < \frac{1}{\binom{m}{2}}$ }
          \Return $H$
        \EndIf
      \EndWhile 
    \EndFunction
  \end{algorithmic}
\end{small}
\end{algorithm}

\begin{algorithm}[t]
\begin{small}
  \caption{Sampling $k$-subgraph prop.\ to its degree in $G^{(k)}$}
  \label{alg:prop}
  \begin{algorithmic}[1]
    \Require Graph $G = (V,E)$, subgraph size $k$, error $\varepsilon>0$
    \Ensure $v \subgraph{k} G$ sampled from $V^{(k)}$ with probability proportional to its degree
    \Function{DegreePropSampling}{$G,k,\varepsilon$}
      \If{k = 2}
        \Return an edge $(u,v) \in E$ with probability  proportional to $(d(u) + d(v) - 2)$
      \EndIf
      \State $v_c \gets$ \Call{Uniform\-Sampling}{$G,k,\varepsilon$}
      \State $d_c \gets$ degree of $v_c$
      \For{$\ceil{t_k(\varepsilon)}$ times}
        \If{$\mathit{random}(0,1) < \frac{1}{2}$}
          \State $v_n \gets$ \Call{Uniform\-Sampling}{$G,k,\varepsilon$}
          \State $d_n \gets$ degree of $v_n$
          \If{$\mathit{random}(0,1) < \frac{d_n}{d_c}$}
            \State $v_c \gets v_n$, $d_c \gets d_n$
          \EndIf
        \EndIf
      \EndFor
      \Return $v_c$
    \EndFunction
  \end{algorithmic}
\end{small}
\end{algorithm}

The key observation is that sampling a 2-subgraph (edge) can be done in 2 steps: 
(1) sampling a node in $G$ with probability proportional to its degree; and
(2) sampling uniformly an adjacent edge.
This approach can be generalized to any $k>2$ as follows:
\squishlist
  \item[(1)] sample a node $v$ in $G^{(k-1)}$ with probability  proportional to its degree;
  \item[(2)] sample uniformly at random an edge adjacent to $v$; 
  denote this edge by $(v,u) \in E^{(k-1)}$.
  \item[(3)] output a $k$-subgraph $H$ whose node set is the union of nodes of $v$ and $u$ with appropriate probability.
\squishend
In the proposed \rss approach, step (1) is performed in $\Call{Degree\-Prop\-Sampl\-ing}{G,k-1,\varepsilon}$,
while steps (2) and (3) are performed in $\Call{Uniform\-Sampl\-ing}{G,k,\varepsilon}$.
We now discuss these two functions in more detail.

\block{DegreePropSampling.}
To sample a node $v$ in~$V^{(k)}$ 
with probability proportional to its degree in the state graph $G^{(k)}$,
we apply the \mh algorithm on a \emph{complete graph}.
Let us assume we can sample $v_c$ in  $V^{(k)}$ uniformly at random, 
which is done by $\text{\sc Uniform}\-\Call{Sampl\-ing}{G,k,\varepsilon}$ as explained later. 
Starting from $v_c$ we then sample a next state $v_n \in V^{(k)}$ uniformly at random. 
This is regarded as a random walk on the \emph{complete graph} with nodes $V^{(k)}$. 
Since those samples are uniform, the stationary distribution is uniform, $\pi_x \sim 1$, 
and needs to be converted into $\pi_x \sim d(x)$ for any $x \in V^{(k)}$. 
Hence, we calculate the degrees $d(v_c)$ and $d(v_n)$, and 
accept $v_n$ as a new node $v_c$ with probability $\min \left\{1,\frac{d(v_n)}{d(v_c)}\right\}$. 
If we continue this walk for more than $t_{k}(\varepsilon)$ steps, 
$v_c$ becomes an approximate sample of $V^{(k)}$ with probability proportional to its degree.

We calculate an upper bound on the mixing time $t_{k}(\varepsilon)$ of 
$\Call{Degree\-Prop\-Sampling}{G,k,\varepsilon}$ by applying again the canonical-paths argument.
A crucial element of the construction is that the underlying graph of the Markov chain is the 
complete graph with $|V^{(k)}|$ nodes.
The target stationary distribution is $\pi_v = \frac{d(v)}{Z}$, where 
$Z = \sum_{v \in V^{(k)}} d(v) = 2|E^{(k)}|\leq k\Delta|V^{(k)}|$, and $Z \geq |V^{(k)}|$.
The transition probability from a node $u$ to $v$ is 
$p(u,v) = \frac{1}{2|V^{(k)}|} \min \left\{1,\frac{d(v)}{d(u)}\right\}$, and 
$Q(u,v) =  \pi_u p(u,v) =\frac{d(u)}{Z}\frac{1}{2|V^{(k)}|}\min \left\{1,\frac{d(v)}{d(u)}\right\} =\frac{\min\{d(v),d(u)\}}{2Z|V^{(k)}|}$.
The quantity $\overline\rho$ is calculated as follows:
\begin{small}
\begin{eqnarray*}
\overline\rho 
  & = & \max_{(u,v) \in V^{(k)} \times V^{(k)}} \frac{1}{Q(u,v)} \sum_{\gamma_{xy} \in \Gamma \wedge \gamma \ni (u,v)} \pi_x \pi_y |\gamma_{xy}|  \\
  & = & \max_{(u,v) \in V^{(k)} \times V^{(k)}} \frac{2Z|V^{(k)}|}{\min\{d(v),d(u)\}} \frac{d(u)}{Z} \frac{d(v)}{Z} \\
  & = & 2\frac{|V^{(k)}|}{2|E^{(k)}|} \max_{u \in V^{(k)}} d(u) \\
  &\leq& 2k\Delta.
\end{eqnarray*}
\end{small}
A bound on mixing time $t_k(\varepsilon)$ is obtained by
\begin{small}
\begin{eqnarray*}
t_k(\varepsilon)
  &\leq& \max_{x\in V^{(k)}}  \overline\rho \left(\ln \left( \frac{Z}{d(x)} \right) + \ln \varepsilon^{-1}\right) \\
  &\leq& \overline\rho(\ln (k\Delta|V^{(k)}|) + \ln \varepsilon^{-1}) \\
  &\leq& 2k\Delta (\ln k + \ln \Delta + k \ln |V| + \ln \varepsilon^{-1}) \\
  & = & \bigO(k^2 \Delta \ln |V|).
\end{eqnarray*}
\end{small}
The total cost of $\Call{Degree\-Prop\-Sampling}{G,k,\varepsilon}$ is 
$D_k = \bigO(k^2 \Delta \ln|V| U_k)$, 
where $U_k$ is the total computational cost of $\Call{Uniform\-Sampling}{G,k,\varepsilon}$.

Note that $\Call{Degree\-Prop\-Sampling}{G,2,\varepsilon}$ runs in constant time $\bigO(1)$, 
by pre-computing $(d(u)+d(v)-2)$ for each edge $(u,v) \in E$, 
which is the degree of 2-subgraph with nodes $\{u,v\}$, on $G^{(2)}$.

\block{UniformSampling.}
We now discuss how to sample a $k$-subgraph uniformly at random.
We first call $\text{\sc Degree\-Prop}\-\Call{Sampling}{G,k-1,\varepsilon}$ and obtain $v$ in $V^{(k-1)}$ 
with probability proportional to its degree. 
Then we sample a neighbor state $u$ of $v$, uniformly at random,
and then we obtain the subgraph $H$ whose nodes are the union of nodes in $v$ and $u$. 
When $k=2$, the subgraph $H$ is an uniform sample among all 2-subgraphs (edges). 
When $k>2$, however, the number of edges in $E^{(k-1)}$ that outputs this same $H$ 
is equal to the number of edges among $\{F \subgraph{k-1} H\}$; 
let $m$ be the number of such~$F$. 
The subgraph $H$ is accepted with probability $1/\binom{m}{2}$.
If accepted, $H$ is the output subgraph.
If rejected, we repeat until some subgraph is accepted. 
Note that $m$ is at most $k$.
Hence, on expectation we have to repeat the process  
$\binom{m}{2} \leq \binom{k}{2} = \bigO(k^2)$ times.

Thus, the computational complexity of $\text{\sc Uniform}\-\Call{Sampling}{G,k,\varepsilon}$ 
subroutine is 
$U_k = \bigO(k^2 D_{k-1})$, 
where $D_{k-1}$ is the computational cost of $\text{\sc Degree\-Prop}\-\Call{Sampling}{G,k-1,\varepsilon}$.
Unrolling the recurrences we obtain the overall complexity of \rss,
\begin{small}
\begin{eqnarray*}
  U_k 
  & = & \bigO(c^{k-3}(k!)^2 ((k-1)!)^2 \Delta^{k-3} (\ln|V|)^{k-3} )\,,
\end{eqnarray*}
\end{small}
where $c$ is a constant independent of $k$, $\Delta$, $|V|$ and the other variables. 
We obtain the following theorem. 
\begin{theorem}[Computational cost of \rss]
  \label{thm:rss}
  \rss takes time $\bigO(c^{k-3}(k!)^2 ((k-1)!)^2 \Delta^{k-3} (\ln|V|)^{k-3})$.
\end{theorem}

We note that \rss is significantly more efficient than \mcmcsampling. 
Considering $k$ to be small, and ignoring exponentials and factorials in $k$,
the prohibitive factor $\Delta^{k+1}D|V| \ln |V|$ in \mcmcsampling
has given its place to the mild factor $(\Delta \ln|V|)^{k-3} $ in \rss.

\begin{algorithm}[t]
\begin{small}
  \caption{Sampling $k$-subgraph prop.\ to its degree in $G^{(k)}$}
  \label{alg:prop2}
  \begin{algorithmic}[1]
    \Require Graph $G = (V,E)$, subgraph size $k$, error $\varepsilon>0$
    \Ensure $H \subgraph{k} G$ sampled from $V^{(k)}$ with probability proportional to its degree
    \Function{DegreePropSampling+}{$G,k,\varepsilon$}
      \If{k = 2}
        \Return an edge $(u,v) \in E$ with probability proportional to $(d(u) + d(v) - 2)$
      \EndIf
      \State $v_c \gets$ \Call{DegreePropSampling+}{$G,k-1,\varepsilon$}
      \State $u_c \gets$ uniformly sampled neighbor of $v_c$
      \State $H_c \gets$ $k$-subgraph with nodes in $v_c$ and $u_c$
      \State $d_c \gets$ degree of $H_c$ on $G^{(k)}$
      \State $m_c \gets $ number of $(k-1)$-subgraphs of $H_c$
      \State $f_c \gets d_c / \binom{m_c}{2}$
      \For{$\ceil{t'_k(\varepsilon)}$ times}
        \If{$\mathit{random}(0,1) < \frac{1}{2}$}
          \Continue
        \EndIf
        \State $v_n \gets$ \Call{DegreePropSampling+}{$G,k-1,\varepsilon$}
        \State $u_n \gets$  uniformly sampled neighbor of $v$
        \State $H_n \gets$ $k$-subgraph with nodes in $v_n$ and $u_n$
        \State $d_n \gets$ degree of $H_n$ on $G^{(k)}$
        \State $m_n \gets $ number of $(k-1)$-subgraphs of $H_n$
        \State $f_n \gets d_n / \binom{m_n}{2}$
        \If{$ \mathit{random}(0,1) < \frac{f_n}{f_c}$}
          \State $H_c \gets H_n$, $f_c \gets f_n$
        \EndIf
      \EndFor
      \Return $H_c$
    \EndFunction
  \end{algorithmic}
\end{small}
\end{algorithm}

\subsection{\rssopt: an improved variant of \rss}

A source of computational inefficinecy for the \rss scheme
is that \textsc{UniformSampling} may reject a large number of samples.
To address this issue, 
we can incorporate the rejection probability into the proposal step of the \mh algorithm
in \textsc{DegreePropSampling}. 
The revised \textsc{DegreePropSampling} is shown in \algref{prop2} as \textsc{DegreePropSampling+}.
Note that there are no recursive calls to \textsc{UniformSampling} anymore.

\Call{Degree\-Prop\-Sampling+}{$G,k,\varepsilon$} performs the edge-sampling process that is done in 
\Call{Uniform\-Sampling}{$G,k,\varepsilon$} without rejection. 
First, it samples an edge $(v_c,u_c) \in E^{(k-1)}$ uniformly at random using 
\Call{Degree\-Prop\-Sampling+}{$G,k-1,\varepsilon$}. 
Let $H_c \in V^{(k)}$ be $k$-subgraph whose node set is the union of nodes in $v_c$ and $u_c$. 
Since, $(v_c, u_c)$ is sampled uniformly, 
this particular $H_c$ appears with probability proportional to $\binom{m_c}{2}$, 
where $m_c$ is the number of $(k-1)$-subgraphs of $H_c$. 
We need to convert this probability into the one proportional to $d(H_c)$, 
which is the degree of $H_c$ in $G^{(k)}$.
Again, we apply the MH technique. 
Let $f(H)$ be ${d(H)}/{\binom{m(H)}{2}}$, $m(H)$ is the number of $(k-1)$-subgraphs of $H \in V^{(k)}$.
Starting with an edge $(v_c, u_c)\in E^{(k-1)}$ and the corresponding $k$-subgraph $H_c \in V^{(k)}$, 
\Call{Degree\-Prop\-Sampling+}{$G,k,\varepsilon$} samples another $(v_n,u_n) \in E^{(k-1)}$, and 
corresponding $H_n \in V^{(k)}$. 
It accepts $H_n$ as new $H_c$ with probability $\min(1,{f(H_n)}/{f(H_c)})$.
After repeating this walk at least $t'_k(\varepsilon)$ times, $H_c$ becomes an approximate sample of $k$-subgraph proportional to its degree.

The mixing time of \Call{DegreePropSampling+}{$G,k,\varepsilon$} is given by the following lemma.

\begin{lemma}[Mixing time of \rssopt]
\label{lem:rssopt}
The mixing time $t'_k(\varepsilon)$ of \Call{DegreePropSampling+}{$G,k,\varepsilon$} 
is $2k\Delta(k\ln|V| + 3\ln k +\ln \Delta + \ln \varepsilon^{-1}) = \bigO(k^2\Delta \ln|V|)$.
\end{lemma}

\noindent
The overall complexity is obtained by the following theorem based on the lemma above.



\begin{theorem}[Computational cost of \rssopt]
\label{thm:rssopt}
\rssopt takes time $\bigO( c^{k-3} k^2 ((k-1)!)^2 \Delta^{k-3} (\ln |V|)^{k-3})$.
\end{theorem}

\subsection{Analysis of \psrw}

Another sampling method is \psrw~\cite{psrw}. 
The idea of \psrw is similar to \rss 
but instead of $\Call{Degree\-Prop\-Sampling}{G,k,\varepsilon}$, 
it adopts a standard random walk on the $(k-1)$-state graph, 
$G^{(k-1)}$ and obtains a node of $V^{(k-1)}$ with probability pro\-por\-tional to its degree.
The authors of \psrw do not provide the mixing time of its random walk, and overall computational costs.
To compare the com\-pu\-ta\-tional cost with \mcmcsampling, \rss, and \rssopt, we obtain the following bound of the mixing time and the com\-pu\-ta\-tional cost.

\begin{lemma}[Mixing time of \psrw]
\label{lem:psrw}
The mixing time of \psrw is $\frac{1}{2} (k-1)!(k-1) \Delta^k (D+k-2) |V|  ((k-1)\ln |V| + \ln (k-1) +\ln \Delta + \ln\varepsilon^{-1}) = \bigO((k-1)!(k-1)^2 \Delta^k (D + k) |V| \ln|V|)$.
\end{lemma}

\begin{table*}[t]
\begin{center}
  \caption{\label{tab:complexity}Computational cost comparison.}
  \begin{tabular}{lll} \toprule
    Method    &   Time complexity & Suppressing $k$ and logarithmic terms \\ \midrule
    \mcmcsampling & $\bigO(k!k^5\Delta^{k+1} (D+k)|V| \ln |V|)$                     &  $\tbigO ( \Delta^{k+1} D |V| )$ \\
    \psrw         & $\bigO(k!k(k-1)^6 \Delta^{k+1} (D + k) |V| \ln|V|)$             &  $\tbigO ( \Delta^{k+1} D |V| )$ \\
    \rss          & $\bigO(c^{k-3}(k!)^2 ((k-1)!)^2 \Delta^{k-3}  (\ln|V|)^{k-3} )$ &  $\tbigO ( \Delta^{k-3} )$ \\
    \rssopt       & $\bigO( c^{k-3} k^2 ((k-1)!)^2 \Delta^{k-3} (\ln |V|)^{k-3})$   &  $\tbigO ( \Delta^{k-3} )$ \\
    \bottomrule
  \end{tabular}
\end{center}
\end{table*}

\begin{theorem}[Computational cost of \psrw]
\label{thm:psrw}
\psrw takes time $\bigO(k!k(k-1)^6 \Delta^{k+1} (D + k) |V| \ln|V|)$.
\end{theorem}

\subsection{Computational cost comparison}

The computational costs of the methods considered in this paper are shown in \tabref{complexity}. 
The variables $\Delta$ and $D$ are the maximum node degree, and the diameter of the input graph, respectively. 
On the right-most column we show the computational costs, considering $k$ as a fixed small constant, and 
suppressing logarithmic terms by $\tbigO(\cdot)$.
Methods \mcmcsampling and \psrw contain terms $|V|$ and $\Delta^{k+1}$ in their computational cost, 
which make them inefficient. 
On the other hand, methods \rss and \rssopt are not directly affected by $|V|$, 
and costs are only proportional to $\tbigO(\Delta^{k-3})$, considering $k$ fixed. 
Thus, \rss and \rssopt are superior to \mcmcsampling and \psrw.

Note that these theoretical computational costs are derived based on the worst-case bounds for each Markov chain. 
The actual and practical costs might be smaller.

\section{Experimental evaluation}
\label{section:experiments}

We conduct experiments to evaluate and compare all methods, 
\mcmcsampling, \psrw, \rss and \rssopt. 
We implement each algorithm in Python 3.5 with libraries NetworkX 2.3 and NumPy 1.16.4.
The basic implementations of \rss and \rssopt are available online.\footnote{https://github.com/ryutamatsuno/rss}
The experiments are conducted on a workstation with 16 Intel Xeon CPU E5-2670 2.60GHz processors and 256 GB RAM memory. 

It should be noted that we choose very small graphs for the experiments, as 
(i) we materialize $G^{(k)}$ for val\-i\-da\-tion purposes, and 
(ii) we run the methods to their theoretical limits.
We observe, however, that in practice, 
the methods converge much faster than the the\-o\-ret\-i\-cal bounds, 
and thus, one could run the methods for a smaller number of steps,
and obtain high-quality samples. 

In Appendix~\ref{section:additional-experiments} we present an experiment with a graph of 1\,million nodes.
We also present two additional experiments:
the sampling times of \rss and \rssopt with higher $k$, and a use case with a real-world graph.

\subsection{Uniformity of \rss and \rssopt}
We check 
whether RSS and \rssopt give truly uniform samples.
We also check how fast the chain mixes in practice.

\block{Setting.}
Given a graph $G=(V,E)$, we enumerate all possible $k$-subgraphs. 
Then we obtain $N_{s} = 1000|V^{(k)}|$ samples using \rss and \rssopt. 
We calculate the error of the output distribution among the obtained samples. 
The evaluation is based on the loss used in the definition of the mixing time~\cite{mix},
\begin{align}
\begin{small}
  \label{eq:L}
  \mathit{Loss} = \frac{1}{2}\sum_{v \in V^{(k)}} \left| \frac{N_v}{N_{s}} - \frac{1}{|V^{(k)}|} \right|,
\end{small}
\end{align}
\noindent
where $N_{s}$ is the total number of samples, $N_v$ is the number of samples of a subgraph $v$ obtained by each algorithm. The term $\frac{1}{|V^{(k)}|}$ represents the uniform probability for all subgraphs. From the definition of the mixing time, $\mathit{Loss}$ is smaller or equal than the error $\varepsilon$.

We set $\varepsilon$ to $0.05$, and set $k$ to 3 and 4. We run this experiment 10 times for each $k$, and report the averages and the standard deviations.

\block{Dataset.}
We use Zachary's karate club \cite{Zachary} as the input graph $G=(V,E)$. The number of nodes, $|V|$, and edges, $|E|$, are 34 and 78, respectively. The number of $3$-subgraphs is $|V^{(3)}|= 438$, and the number of 4-subgraphs is 
$|V^{(4)}|= 2\,363$.

\block{Results.} The results, shown in \tabref{uniformity}, show the average loss and standard deviation over 10 runs.
Loss is smaller than $\varepsilon$ (0.05) and standard deviation is small, which shows that \rss and \rssopt give uniform samples as theoretically shown in the previous sections. 

\begin{table}[t]
  \caption{Result of uniformity experiment of \rss and \rssopt.}
  \label{tab:uniformity}
  \centering
  \begin{tabular}{lcc}
    \toprule
    & \multicolumn{2}{c}{Loss}  \\ \midrule
    $k$   &  \rss & \rssopt \\ \midrule
    3 & 0.0130$\pm$0.0003 & 0.0128$\pm$0.0005  \\
    4 & 0.0126$\pm$0.0001 & 0.0126$\pm$0.0001  \\
    \bottomrule
  \end{tabular}
\end{table}

Next, we vary the number of steps that we perform before sampling.
We set the number of steps of \rss and \rssopt to smaller values than the theoretical bound. 
The ratio of the number of steps to the theoretical bound is varied from $0, 10^{-3}$ to $10^0=1$. 

\begin{figure}[t]
  \centering
  \includegraphics[width=0.5\linewidth]{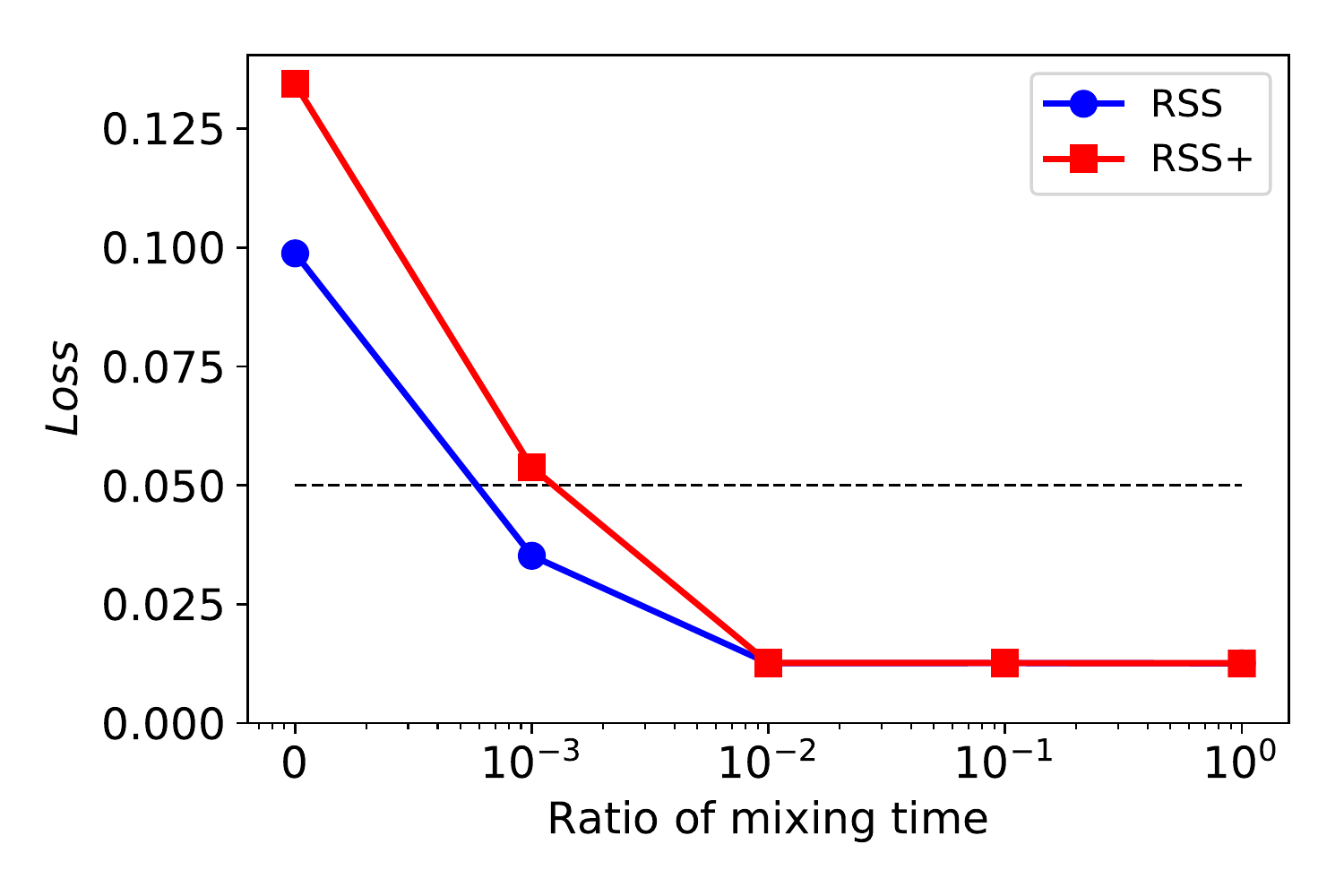}
  \caption{Loss of \rss and \rssopt as a function of the ratio of the theoretical mixing time.}
  \label{fig:uni}
\end{figure}

The result is shown in \figref{uni}.
It shows the loss as a function of the ratio of the number of steps to the theoretical bound.  
The dashed black line indicates when $Loss = \varepsilon = 0.05$; 
when loss becomes smaller than $\varepsilon$ we regard the output as uniform.
Without any random-walk steps, i.e., ratio $0$, 
the outputs of \rss and \rssopt are not uniform, as expected. 
The loss converges to 0.0126 at around ratio $10^{-2}$. 
These results show that, in practice, we may perform a much smaller number of random-walk steps 
than the theoretical mixing-time bounds, and still get useful results.

\begin{figure*}[t]
  \centering
  \subfigure[$k=3$]{
    \includegraphics[width=0.315\linewidth]{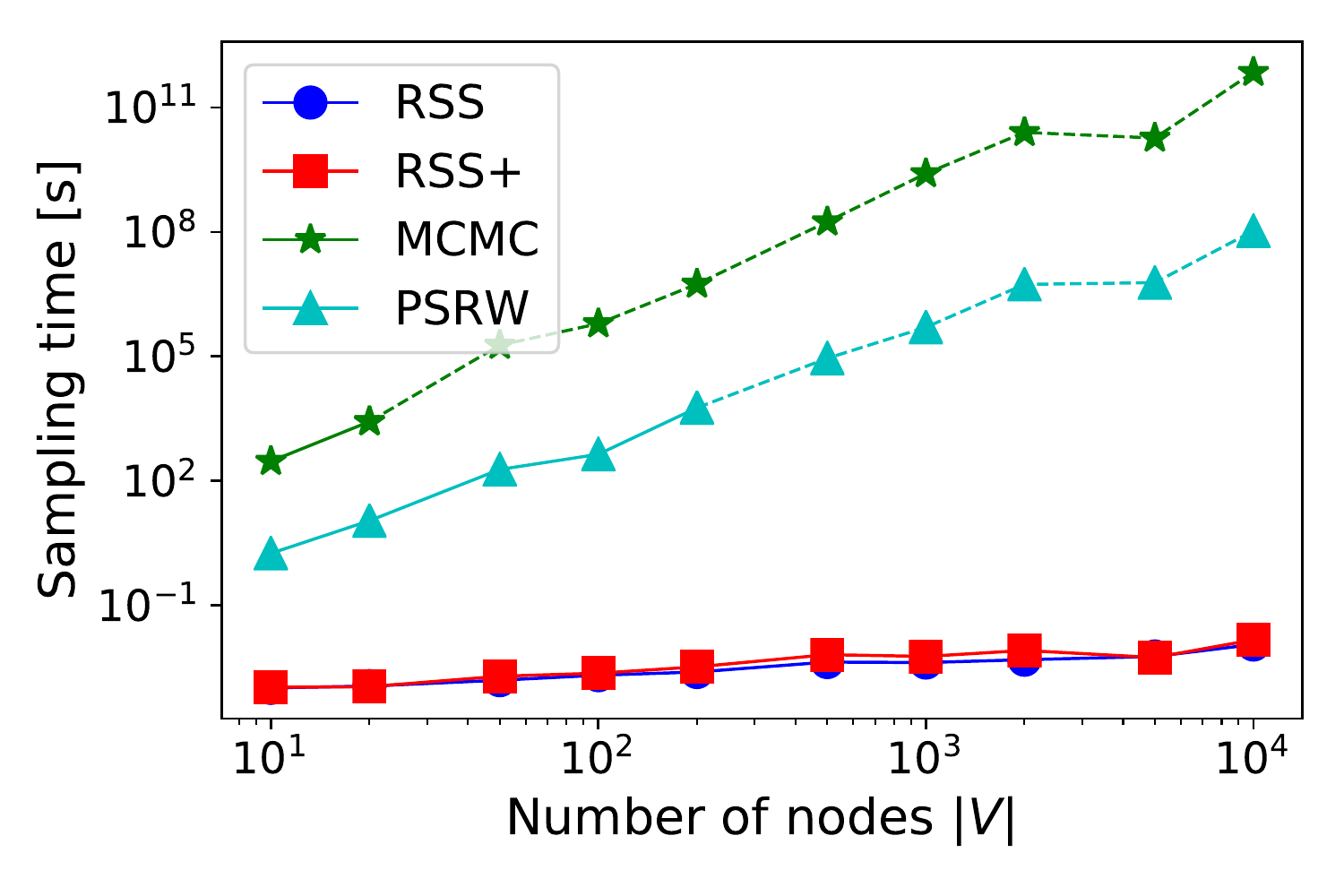}
    \label{fig:samplingtime_k3}
  }%
  \subfigure[$k=4$]{
    \includegraphics[width=0.315\linewidth]{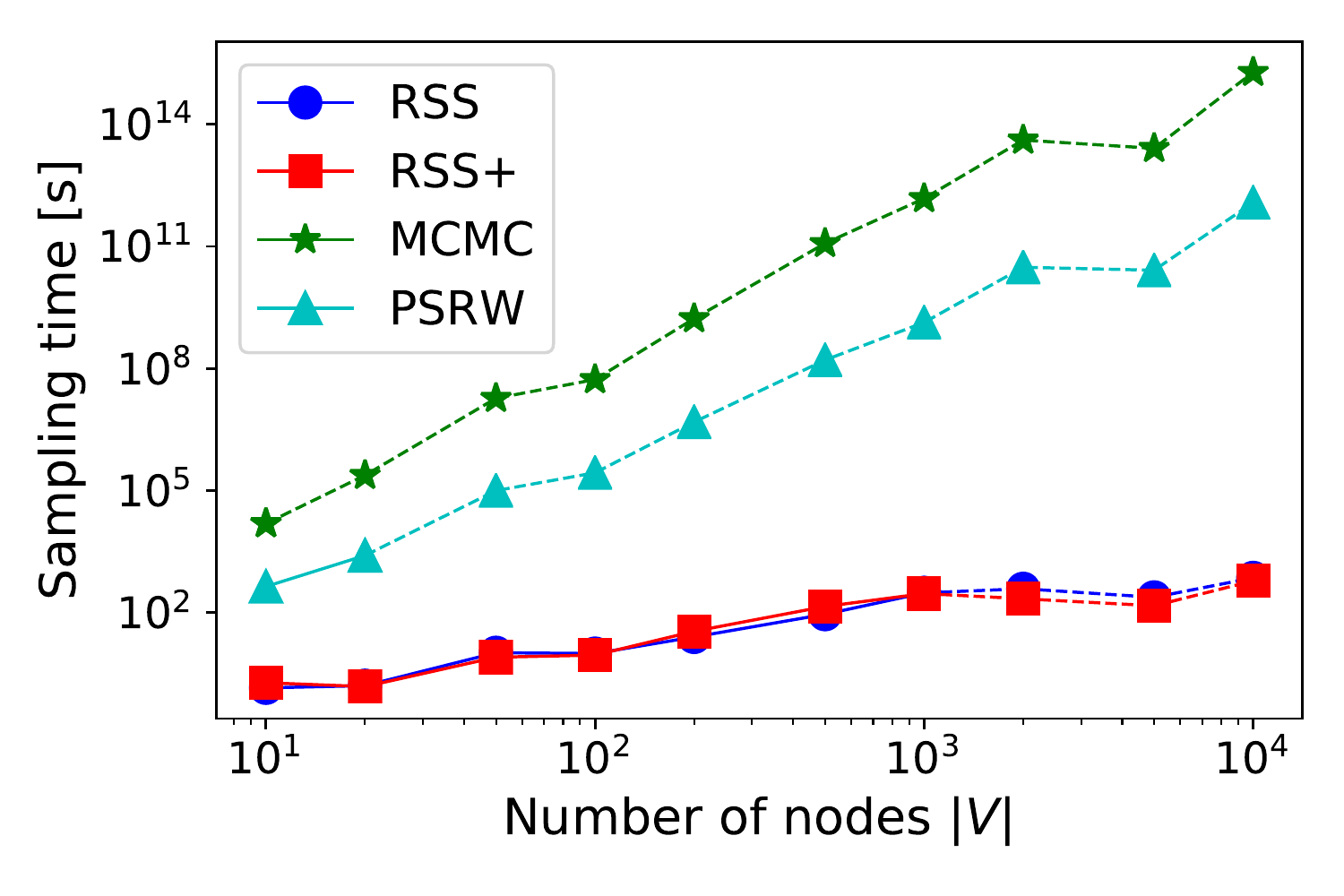}
    \label{fig:samplingtime_k4}
  }%
  \subfigure[$k=5$]{
    \includegraphics[width=0.315\linewidth]{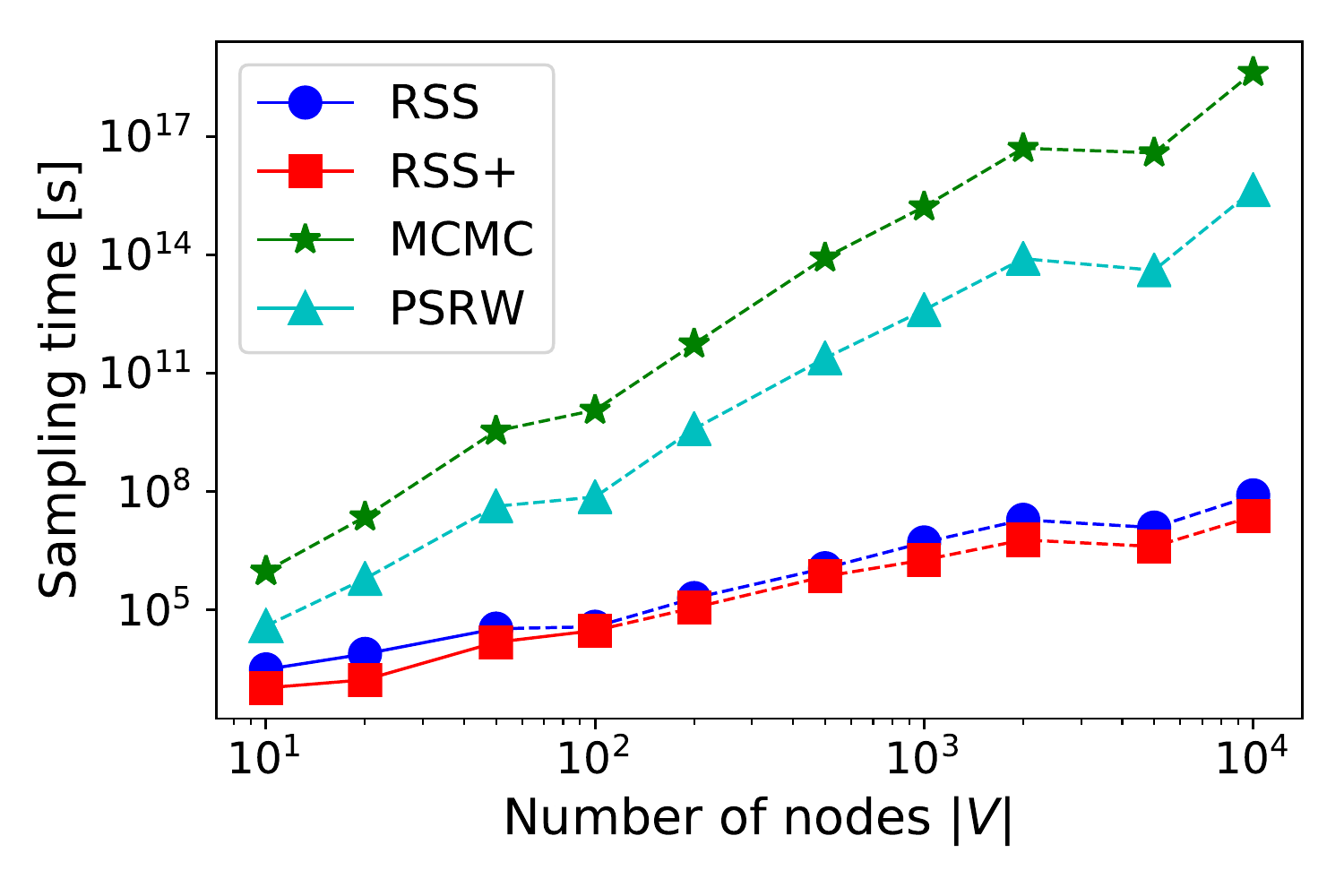}
    \label{fig:samplingtime_k5}
  }
  \caption{Sampling Time (seconds) of \mcmcsampling, \psrw, \rss and \rssopt for different size of BA graphs. Solid line shows actual sampling time, and dashed lines represents estimated time.}
  \label{fig:samplingtime}
\end{figure*}

\subsection{Sampling time comparison with BA graphs}

We compare the sampling time, i.e., time to obtain one $k$-subgraph from the input graph, for \mcmcsampling, \psrw, \rss, and \rssopt. 

\block{Setting.}
We compare sampling times of each method for different size of graphs and different $k$. We fix the error $\varepsilon$ to $0.05$ and vary the size of the input graphs from $10^1$ to $10^4$. We also vary $k$ from 3 to 5. We measure the time to obtain one $k$-subgraph for each method for each setting 10 times, and report the averages. For small graphs and small $k$, we measure the actual time. For big graphs and for big $k$, we report the estimated time,
 based on the time taken to make 100 steps.

\block{Dataset.}
We use Barab\'asi-Albert (BA) model \cite{BAmodel}, which is a well-studied graph generating model of random scale-free networks using preferential attachment. The BA model parameter $m$ is set to 2. 
We vary $|V|$ from $10,20,50,...$ up to $10\,000$.

\block{Results.}
The results are shown in \figref{samplingtime}. Solid lines represent actual times and dashed lines represent estimated times. 
%
%
%
\mcmcsampling and \psrw run slowly, e.g., for $|V| = 100, k=5$ \mcmcsampling takes $10^{10}$ seconds and \psrw takes $10^7$ seconds on the estimation.

\rss and \rssopt are both quite fast. For $k=3$, they run in almost constant time no matter how the size of the graph is big. Indeed, the theoretical running time of both \rss and \rssopt is $\bigO(1)$. 
For $k=4$ and $k=5$, the sampling time increases with $|V|$, however, 
compared to \mcmcsampling and \psrw, the speed of the increase is mild. 
In addition, we confirm that \rssopt is faster than \rss, 
validating our theoretical results.
The detailed comparison between \rss and \rssopt with higher $k$ can be found in 
Appendix~\ref{section:additional-experiments}


\section{Conclusion}
\label{section:conclusion}

In this paper, we have studied the problem of sampling $k$-subgraphs from a given graph.
We have analyzed \mcmcsampling, the standard \mcmc approach for this problem, and \psrw, a state-of-the-art \mcmc method. We improved the upper bounds for the mix\-ing times for both methods, 
using the canonical-paths technique. 
In addition, we have proposed novel \mcmc methods, \rss and \rssopt, 
which sample $k$-sub\-graphs by sampling $(k-1)$-sub\-graphs in a recursive manner.
We have derived the theoretical mixing time and the com\-pu\-ta\-tional costs for the proposed methods. 
We per\-formed experiments to compare \rss and \rssopt with the ex\-ist\-ing methods. 
We validated that \rss and \rssopt give uni\-form samples, and 
they are significantly faster than the existing methods. 

{
\bibliographystyle{siam}
\bibliography{reference}
}

\begin{appendix}

\section*{\Large Appendix}
\medskip

\section{Proofs}
\label{section:proofs}

\subsection{Lemma~\ref{lem:diameter}: Diameter of $k$-state graph~$G^{(k)}$} 
The diameter of the $k$-state graph $G^{(k)}$ is at most $(D+k-1)$, where $D$ is the diameter of $G$.

\begin{proof}
Consider distinct $x$ and $y$ in $G^{(k)}$, with $x = (V_x,E_x)$ and $y = (V_y,E_y)$.

If $V_x \cap V_y\neq \emptyset$, starting with $V_x$ as a current node set, 
we add one node in $V_y$ that is adjacent to the current node set, and 
remove one node from the current node set except for the nodes that have been added, 
so that the induced subgraph of $G$ with the current node set remains connected.
This step corresponds to a one-step walk on $G^{(k)}$. 
After $(k - |V_x \cap V_y|)$ such steps we obtain a path from $x$ to $y$ on $G^{(k)}$. 
Thus, the length of the shortest path from $x$ to $y$ is at most $(k - |V_x \cap V_y|) \leq k$.

If $V_x \cap V_y = \emptyset$, we consider a shortest path from a node in $V_x$ to a node in $V_y$. 
The length of such a path is at most $D$.
We add and remove nodes in the same manner as above; 
starting with $V_x$ as a current node set, 
we add one node from the path that is adjacent to the current node set, 
and remove one node from the set. 
Once we add all nodes in the path, i.e., 
after at most $D$ steps,  
the current node set contains one node in $V_y$. 
Then we add nodes from $V_y$ and remove one node from the set, as above. 
It takes $k-1$ steps until the set becomes equal to $V_y$. 
The total number of steps is at most $(D+k-1)$, 
which shows that the length of the corresponding walk from $x$ to $y$ is at most $(D+k-1)$.

Hence, for any choice of $x$ and $y$ in $V^{(k)}$, 
their distance on $G^{(k)}$ is at most $(D+k-1)$, and thus, 
the diameter of $G^{(k)}$ is at most $(D+k-1)$.
\hfill\end{proof}

\subsection{Lemma~\ref{lem:rssopt}: Mixing time of \rssopt} 
The mixing time $t'_k(\varepsilon)$ of the algorithm 
\Call{DegreePropSampling+}{$G,k,\varepsilon$}  
is $2k\Delta(k\ln|V| + 3\ln k +\ln \Delta + \ln \varepsilon^{-1}) = \bigO(k^2\Delta \ln|V|)$.

\begin{proof}
We apply the canonical-paths argument to obtain a bound on the quantity $\overline\rho$, 
used in Inequality~(\ref{equation:mixing-rho}), 
for bounding the mixing time of the Markov chain.
The set of states of this Markov chain is $\Omega = E^{(k-1)}$. 
The underlying graph is a complete graph with nodes $E^{(k-1)}$. 
The desired stationary distribution is $\pi_x = \frac{1}{Z}f(x)$, 
where $f(x) = d(v(x))/\binom{m(v(x))}{2}$, 
$v(x)$ is the corresponding node in $V^{(k)}$ whose node set is the union of nodes in $x$,
$d(u)$ is the degree of $u$,
and
$m(u)$ is the number of $(k-1)$-subgraphs in $u$.
We have
\begin{equation*}
Z   =   \sum_{x \in \E^{(k-1)}} f(x)  = \sum_{v \in V^{(k)}} d(v) = 2|E^{(k)}|  \leq  k\Delta|V^{(k)}| .
\end{equation*}
Thus, 
\begin{align*}
p(x,y) = \frac{1}{2|E_{k-1}|} \min \left\{1, \frac{f(y)}{f(x)}\right\},
\end{align*}
and 
\begin{eqnarray*}
Q(x,y) 
  & = & \pi_x p(x,y) \\
  & = & \frac{f(x)}{Z}\frac{1}{2|E_{k-1}|} \min \left\{1, \frac{f(y)}{f(x)}\right\} \\
  & = & \frac{\min \{f(x), {f(y)}\}}{2Z|E_{k-1}|}.
\end{eqnarray*}
We can now bound $\overline\rho$ as follows:
\begin{eqnarray*}
\overline\rho 
  &  = & \max_{(u,v) \in E_{k-1}\times E_{k-1}} \frac{1}{Q(u,v)} \sum_{\gamma_{xy} \ni (u,v)} \pi_x \pi_y |\gamma_{xy}|\\
  &  = & \max_{(u,v) \in E_{k-1}\times E_{k-1}} \frac{2Z|E_{k-1}|}{\min\{f(u),f(v)\}} \frac{f(u)}{Z} \frac{f(v)}{Z}\\
  &  = & 2 \frac{|E_{k-1}|}{Z} \max_{(u,v) \in E_{k-1}\times E_{k-1}}  \max\{f(u),f(v)\} \\
  &  = & 2 \frac{|E_{k-1}|}{2|E_{k}|} \max_{u \in E_{k-1}}  f(u) \\
  &\leq& 2 \max_{u \in V^{(k)}}  d(u) \\
  &\leq& 2 k \Delta,
\end{eqnarray*}
where we use the fact that $\frac{|E_{k-1}|}{2|E^{(k)}|} \leq 1$, when $k \ll |V|$. 
Hence, by Inequality~(\ref{equation:mixing-rho}), a bound on the mixing time $t'_k(\varepsilon)$ can be obtained as follows:
\begin{eqnarray*}
t'_k(\varepsilon)
  &  = & \max_{x\in V^{(k)}} \tau_x(\varepsilon) \notag\\
  &  = & \max_{x\in V^{(k)}} \overline \rho (\ln {\pi_x}^{-1} + \ln \varepsilon^{-1}) \notag \\
  &  \leq & 2 k\Delta \max_{x\in V^{(k)}}  (\ln \frac{Z}{f(x)} + \ln \varepsilon^{-1}) \notag \\
  &  \leq & 2 k\Delta \max_{x\in V^{(k)}}  (\ln \frac{k\Delta|V^{(k)}| \binom{m(v(x))}{2}}{d(v(x))} + \ln \varepsilon^{-1}) \notag \\
  &\leq& 2 k\Delta (\ln (k^3\Delta |V^{(k)}|) + \ln \varepsilon^{-1}) \notag \\
  &\leq& 2 k\Delta (3\ln k + \ln \Delta + k \ln |V| + \ln \varepsilon^{-1}) \notag \\
  &  = & \bigO(k^2 \Delta \ln |V|).
\end{eqnarray*}
\hfill\end{proof}

\subsection{Lemma~\ref{lem:psrw}: Mixing time of \psrw}
The mixing time of the algorithm \psrw is $\frac{1}{2} (k-1)!(k-1) \Delta^k (D+k-2) |V|  ((k-1)\ln |V| + \ln (k-1) +\ln \Delta + \ln\varepsilon^{-1}) = \bigO((k-1)!(k-1)^2 \Delta^k (D + k) |V| \ln|V|)$.

\begin{proof}
We apply the canonical-paths technique to upper bound the mixing time. 
We consider a random walk on $G^{(k-1)}$. 
Note that we add a self-loop to each node with probability $\frac{1}{2}$ 
to avoid periodicity.
The state space of the corresponding Markov chain is $V^{(k-1)}$.
The stationary distribution is $\pi(x) =\frac{1}{Z} d(x)$, where 
$Z = \sum_{x \in V^{(k-1)}} d(x) = 2|E^{(k-1)}|.$
The transition probability is 
$p(u,v) = \frac{1}{2d(u)} \geq \frac{1}{2\Delta_{k-1}},$
and
$Q(u,v) = \pi(u)p(u,v) = \frac{1}{2Z}$.
We choose as canonical path for $x$ to $y$ to be one of the shortest paths on $G^{(k-1)}$, 
hence $\gamma_{xy} \leq (D + k - 2)$.
The quality $\overline\rho$ is calculated as follows:
\begin{eqnarray*}
  \overline\rho
  & = & \max_{(u,v) \in E^{(k-1)}} \frac{1}{Q(u,v)} \sum_{\gamma_{xy} \in \Gamma \wedge \gamma_{xy} \ni (u,v)} \pi(x) \pi(y) |\gamma_{xy}| \\
  &\leq & \max_{(u,v) \in E^{(k-1)}} 2Z \sum_{\gamma_{xy} \in \Gamma \wedge \gamma_{xy} \ni (u,v)} \frac{d(x)}{Z} \frac{d(x)}{Z} (D+k-2) \\
  &\leq & 2 \frac{{\Delta_{k-1}}^2}{Z} (D+k-2) 
  \max_{(u,v) \in E^{(k-1)}} |\{\gamma_{xy} \in \Gamma \wedge \gamma_{xy} \ni (u,v)\}| \\
  &\leq & 2 \frac{(k-1)^2\Delta^2 }{2|E^{(k-1)}|}(D+k-2) \left( \frac{|V^{(k-1)}|}{2} \right)^2 \\
  &\leq & \frac{1}{2} (k-1)^2 \Delta^2 (D+k-2)\frac{|V^{(k-1)}|}{2|E^{(k-1)}|} |V^{(k-1)}| \\
  &\leq & \frac{1}{2} (k-1)!(k-1) \Delta^k (D+k-2) |V|.
\end{eqnarray*}

Hence the mixing time $\mixingkpsrw$ of \psrw is bounded as follows:
\begin{eqnarray*}
\mixingkpsrw
&  = & \max_{x\in V^{(k-1)}} \tau_x(\varepsilon) \notag\\
&\leq & \max_{x\in V^{(k-1)}} \overline\rho (\ln \pi(x)^{-1} + \varepsilon^{-1}) \\
&\leq &\overline\rho  (\ln (k-1)\Delta|V^{(k-1)}| + \varepsilon^{-1}) \\
&\leq &\frac{1}{2} (k-1)!(k-1) \Delta^k (D+k-2) |V| 
 (\ln (k-1) +\ln \Delta + (k-1)\ln |V| + \ln\varepsilon^{-1}) \\
&=  &\bigO((k-1)!(k-1)^2 \Delta^k (D + k) |V| \ln|V|)
\end{eqnarray*}
\hfill\end{proof}

\subsection{Theorem~\ref{thm:psrw}: Computational cost of algorithm \psrw} 
The total computation cost of the algorithm \psrw is 
$\bigO(k!k(k-1)^6 \Delta^{k+1} (D + k) |V| \ln|V|)$.

\begin{proof}
As with \mcmcsampling, 
each random-walk step of \psrw takes time $\bigO((k-1)^4\Delta)$. 
Hence, using Lemma~\ref{lem:psrw}, 
the computational cost of the random walk on $G^{(k-1)}$ is 
$\bigO((k-1)!(k-1)^6 \Delta^{k+1} (D + k) |V| \ln|V|)$.

\psrw uses the same acceptance and rejection process as $\Call{UniformSampling}{G,k,\varepsilon}$. 
\psrw takes one edge $(u,v)$ sampled from $E^{(k-1)}$ uniformly at random 
using the random walk on $G^{(k-1)}$. 
It accepts a $k$-subgraph whose node set is a union of nodes in $u$ and $v$ 
with probability $1/\binom{m}{2}$, 
where $m$ is the number of $(k-1)$-subgraphs in that $k$-subgraph, which is at most~$k$. 
Hence,  \psrw performs this step $\bigO(k^2)$ times, in expectation, before it accepts. 
The computational cost of \psrw is $\bigO(k!k(k-1)^6 \Delta^{k+1} (D + k) |V| \ln|V|)$.
\hfill\end{proof}

\section{Additional experiments}
\label{section:additional-experiments}

\subsection{Sampling time difference between \rss and \rssopt}

We run \rss and \rssopt with higher $k$ to validate that \rssopt is faster than \rss.

\block{Setting.} We estimate the sampling times of \rss and \rssopt to obtain one subgraph from the same graph with different $k$. We set $k$ from 3 to 10 and set $\varepsilon$ to 0.05.

\block{Dataset.} We use the same BA graph with 100 nodes, which is the same graph used in the experiment 4.2. The BA model parameter $m$ is set to 2.

\begin{figure}[t]
  \centering
  \includegraphics[width=0.5\linewidth]{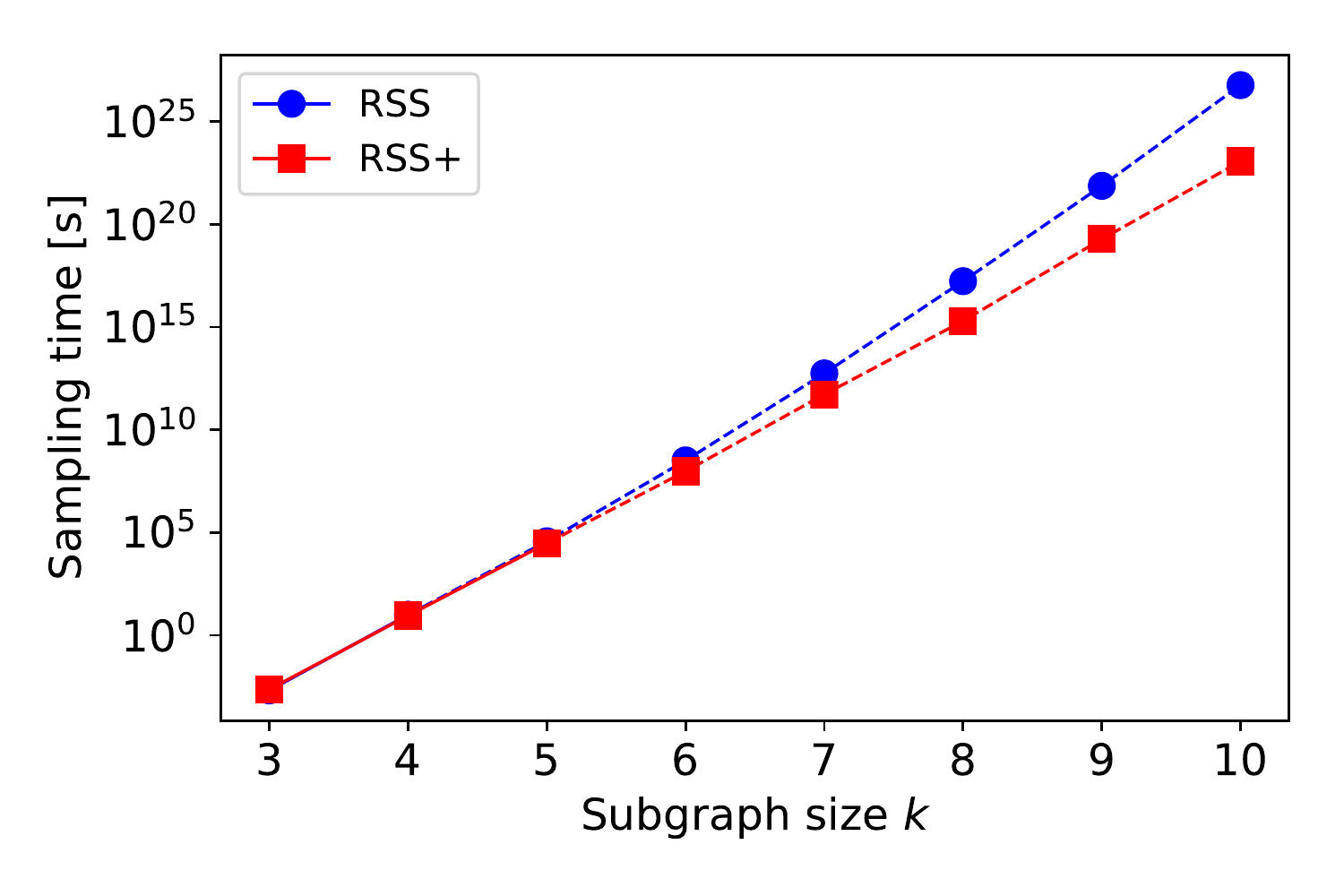}
  \caption{Sampling time of \rss and \rssopt as a function of $k$.}
  \label{fig:diff}
\end{figure}

\block{Results.} The results are shown in \figref{diff}. For $k=3$ and $k=4$, the sampling time of \rss and \rssopt are almost the same. Indeed, the time complexities of \rss and \rssopt are the same, $\bigO(k^2)$ for $k = 3$ and $\bigO(k^2 (k-1)^2 \Delta \ln |V|)$ for $k=4$.
When $k>4$, \rssopt runs faster compared to \rss, and the higher $k$ is, the larger the difference. For example, \rssopt is around 100 times faster than \rss at $k=8$, and 5000 times faster at $k=10$. Thus, we confirm that \rssopt is the fastest method among all methods we compared.


\subsection{Mining patterns of  Bitcoin Alpha web}

As an application, we use \rssopt to mine local patterns in a real-world graph. 

\block{Setting.}
We run \rssopt to obtain 3- and 4-subgraphs and analyze the statistics to find interesting patterns. 
We set $\varepsilon=0.05$.

\block{Dataset.}
We use the Bitcoin Alpha web of trust network \cite{kumar2016edge, kumar2018rev2} 
available online from SNAP.\footnote{https://snap.stanford.edu/data/soc-sign-bitcoin-alpha.html} 
This is a weighted signed directed graph whose nodes represent users of the Bitcoin Alpha platform. 
Edges represent rates of trust among users 
in a scale from $-10$ (total distrust) to $+10$ (total trust). 
The number of nodes, $|V|$, is $3\,783$, and the number of edges, $|E|$, is $24\,186$

\block{Results.}
We run \rssopt on an undirected version of the graph, and after obtaining $k$-subgraphs we consider edge directions and weights. 
We obtain $10^4$ subgraphs, for $k=3$ and $4$. 
For memory efficiency and for speeding \rssopt, we keep $G^{(3)}$ in memory.

The results are shown in \tabref{mining}.
For $k=3$, we investigate the ratio of open triplets, triangles, and balanced triangles 
\cite{balancedtheory_original,balancedtheory}. 
A triangle is regarded as balanced if the number of negative edges among them is even.
In this experiment, we consider that there exists a negative undirected edge between two nodes 
if there exists at least one negative directed edge, 
and if there are no negative directed edges and at least one positive directed edge among two nodes, we consider there exists an positive edge. For $k=4$, we calculate the ratio that the 4-subgraph is line-shaped, i.e., four nodes are connected only by one single path with three edges, and the ratio that the subgraph is a clique.
The triangles in the graph are often balanced, and this shows that there exists some local mechanisms about the rating. It is also interesting that almost 93\% of 4-subgraphs are line-shaped, and only 0.037\% of the 4-subgraphs is a clique, showing that the local interactions among users are not active. 
One can also use our techniques to analyze other interesting local structures, 
for example, considering edge directions and edge weights, 
however, such an analysis is beyond the scope of this paper.

\begin{table}[t]
  \caption{Statistics of patterns of Bitcoin Alpha web}
  \label{tab:mining}
  \centering
  \begin{tabular}{llc}
    \toprule
         & Pattern &  Ratio over all samples\\ \midrule
      $k=3$ 
      &open triplets & 0.97190 \\
      &triangles & 0.02810 \\
      &balanced triangles & 0.02337 \\ \midrule
      $k=4$ 
      &line-shaped    & 0.92798 \\
      &clique & 0.00037 \\
    \bottomrule
  \end{tabular}
\end{table}

\subsection{Motif statistics on a Barab\'asi-Albert graph with 1\,million nodes}
To test the scalability of the proposed method, \rssopt, we test it on a Barab\'asi-Albert (BA) graph with one million nodes.

\block{Setting.}
We run \rssopt on a graph with one million nodes, and obtain 1\,000 samples. 
We set $k$ to 4, and $\varepsilon$ to 0.05. 
We count the frequency of motifs, i.e., small graphs with particular structures, and 
check how the frequency converges with the number of steps in the random walk.

\block{Dataset.}
We generate a BA graph with 1\,million nodes, setting the parameter $m$ to 2, 
thus the number of edges is around 2\,millions. 

\begin{figure}[t]
  \centering
  \subfigure[motif $M_1$]{
    \includegraphics[width=0.12\linewidth]{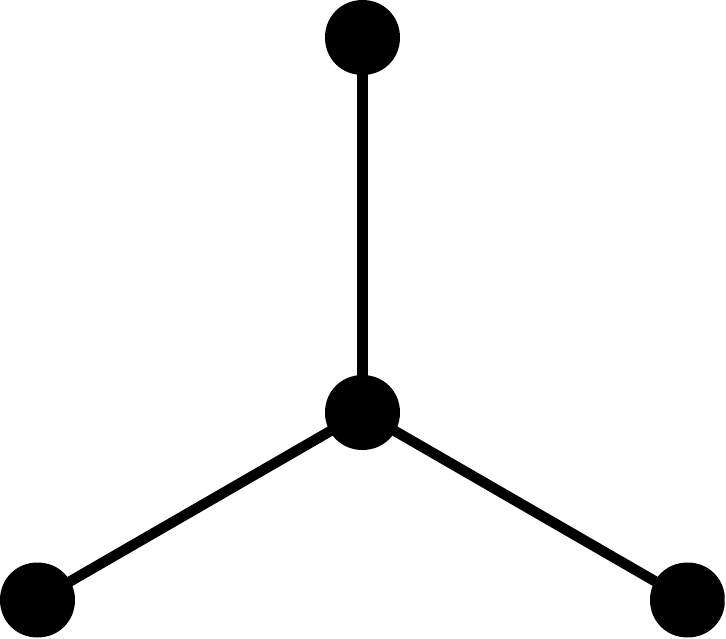}
    \label{fig:motif1}
  }%
  \hspace{0.1\linewidth}
  \subfigure[motif $M_2$]{
    \includegraphics[width=0.12\linewidth]{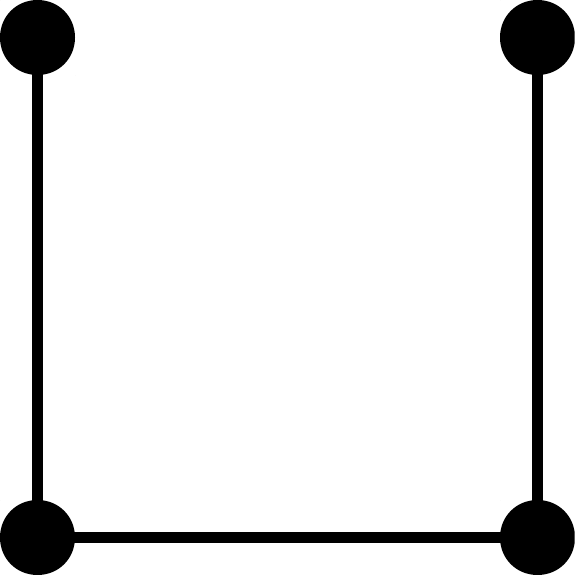}
    \label{fig:motif2}
  }%
  \caption{The two most frequent motifs sampled in a Barab\'asi-Albert (BA) graph. 
  The other four motifs have very low frequencies (less than 1\%).
  }
  \label{fig:motifs}
\end{figure}

\block{Results.}
There are six 4-node motifs, however, only two motifs appear in the vast majority of our samples;
the other four motifs appear very rarely.
This is an effect of the specific structure of the BA graph. 
For instance, we see that our BA graph has very few triangles. 
The two motifs that appear in our samples, $M_1$ and $M_2$, are shown in \figref{motifs}.

\begin{figure}[t]
  \centering
  \includegraphics[width=0.5\linewidth]{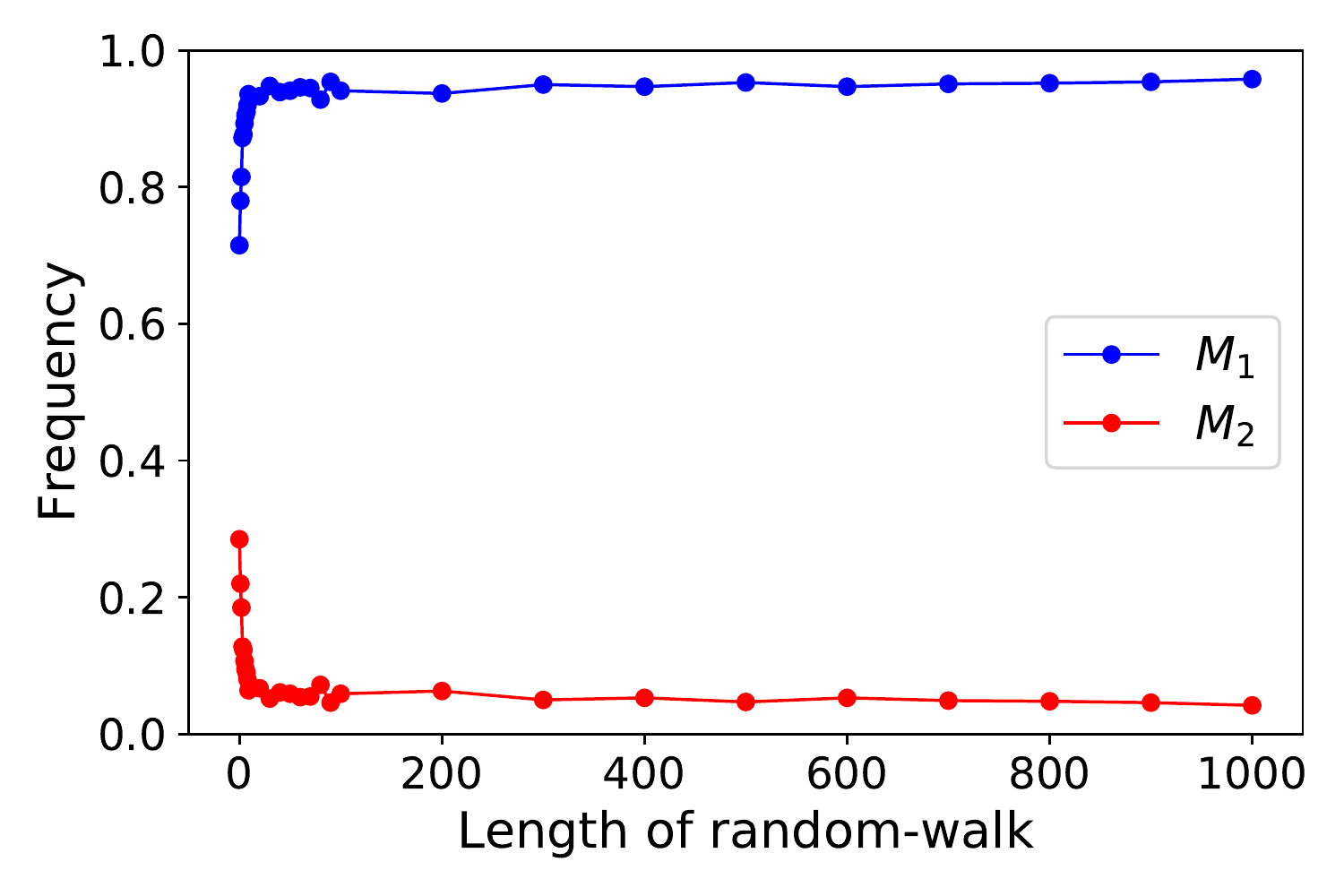}
  \caption{Motif frequencies of $M_1$ and $M_2$ as a function of the length of the random walk. 
  The input graph is a BA graph with 1\,million nodes.}
  \label{fig:motif_frequency}
\end{figure}

In \figref{motif_frequency} we show the frequencies of $M_1$ and $M_2$ 
as a function of the length of random walk. 
It is interesting to observe that the motif frequencies converge with after 10 steps of the random walk, 
while the theoretical bound of the mixing time is $8.6 \times 10^5$. 
Hence, our methods are useful for large graphs by setting appropriate length of random walks, 
which in practice can be much lower than the theoretical upper bounds. 

\end{appendix}

\end{document}